# Diamond nanomechanical resonators protected by a phononic band gap

Xinzhu Li, Ignas Lekavicius[#], and Hailin Wang[*]

Department of Physics, University of Oregon, Eugene, OR 97403, USA

**Abstract**

We report the design, fabrication, and characterization of diamond cantilevers attached to a phononic square lattice. We show that the robust protection of mechanical modes by phononic band gaps leads to a three-orders-of-magnitude increase in mechanical Q-factors, with the Q-factors exceeding $10^6$ at frequencies as high as 100 MHz. Temperature dependent studies indicate that the Q-factors obtained at a few K are still limited by the materials loss. The high-Q diamond nanomechanical resonators provide a promising hybrid quantum system for spin-mechanics studies.

Ultracoherent nanomechanical resonators play an important role in applications as well as fundamental studies ranging from sensitive measurement of mass and force, optomechanics, circuit QED, spin-mechanics, and more generally phonon-mediated hybrid quantum systems [1-6]. For most of these applications, a key figure of merit for the mechanical resonator is the $f \cdot Q$ product, where $f$ is the resonance frequency and $Q$ is the quality factor. Since mechanical waves are not subject to scattering loss into vacuum, the primary loss mechanisms for a mechanical resonator in vacuum are the mechanical clamping loss and the materials loss. A phononic band gap, which protects the mechanical mode from the surrounding environment, can reduce the clamping loss to levels below and in many cases negligible compared with the intrinsic materials loss. With the suitable engineering of a phononic band gap, phonon lifetime of a few minutes has been achieved in MHz high-stress silicon nitride nanomechanical resonators[7-9]. Similarly, phonon lifetime of a few seconds has been observed in GHz silicon nanomechanical resonators at temperatures of a few mK[10]. In addition, bulk acoustic wave (BAW) resonators, while featuring a relatively large size, also offer an alternative approach in reducing the clamping loss[11, 12].

Compared with silicon and silicon nitride nanomechanical resonators, the experimental progress in diamond nanomechanical resonators has been relatively modest due to limitations in diamond nanofabrications. Diamond nanomechanical resonators are of special importance because of the excellent optical and spin properties of color centers such as nitrogen vacancy (NV) and silicon vacancy (SiV) centers. These resonators have been used for studies of spin-mechanics as well as optomechanics [3, 6]. For diamond cantilevers and doubly clamped beams, out-of-plane mechanical modes with frequencies below a few MHz can feature Q-factors near $10^6$[13-17]. Clamping losses, however, severely limit Q-factors of higher frequency out-of-plane or compressional in-plane mechanical modes. For mechanical breathing modes in a diamond microdisk, Q-factors are limited to below $10^5$ by the clamping loss of the supporting pedestal[18]. Diamond optomechanical crystals reported thus far are not protected by a surrounding phononic band gap[19]. The corresponding Q-factors are several orders of magnitude below that of silicon optomechanical crystals. Protection of mechanical modes by a phononic band gap has thus far not been realized in diamond phononic structures.

Here, we report the design, fabrication, and characterization of diamond cantilevers embedded in a 2D phononic crystal lattice. We show that the protection by the phononic band gap can result in a three-order-of-magnitude increase in the mechanical Q-factor, leading to $Q>10^6$ for out-of-plane modes at frequencies as high as 100 MHz and an $f \cdot Q$ product exceeding $10^{14}$. Additional temperature dependent studies indicate that the Q-factors observed are still materials-loss limited at temperatures as low as 8 K. Overall, these results demonstrate that the phononic crystal design and fabrication we have used can circumvent the limitations of diamond nanofabrication, enabling robust protection of mechanical modes by a phononic band gap.

A square phononic crystal lattice has been used in our design[20]. As shown in the phononic structure in Figs. 1a and 1b, cantilevers with a width of 4 μm and with various lengths are attached to the squares in the lattice. The squares with cantilevers attached are surrounded by at least four layers of squares that have no cantilevers attached. Figure 1c plots the phononic band structure of the square lattice shown in Figs. 1a and 1b. The band structure is calculated with the finite element COMSOL Multiphysics software package and with diamond Young's modulus $E=$ 1050 GPa, Poisson ratio $\nu= 0.2$, and mass density $\rho= 3539$ kg/m$^3$. Phononic band gaps centered near 30 MHz, 50 MHz, and 95 MHz can be identified from the band structure. In the second gap, there are also two modes that are confined to individual squares.

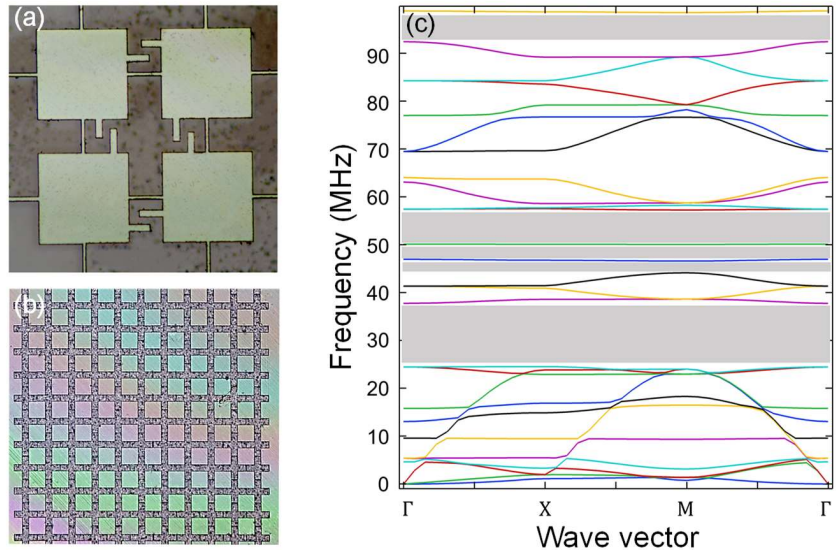

**Figure 1.** (a) Optical image highlighting cantilevers embedded in a square phononic lattice structure. The lengths of the cantilevers range from 11.5 to 15 μm. The period of the lattice is 76 μm. The width and length of the bridge is 1.3 and 20 μm, respectively. (b) Optical image of a section of the phononic structure. The structure is completely released, with a thickness near 2.5 μm. The color fringes in the image reflect thickness variations about 110 nm across one fringe. (c) Calculated phononic band structures of the square lattice with a film thickness of 2.5 μm. The shaded areas highlight phononic band gaps.

The fabrication of the diamond phononic structures follows closely the membrane-in-bulk approach developed in an earlier work, for which a suspended 2D phononic structure is fabricated directly from a bulk diamond film [21]. The fabrication starts with the implantation nitrogen or silicon ions in an electronic grade bulk diamond film about 30 μm in thickness for the creation of color centers near the diamond surface. The phononic structure is fabricated on the front, i.e., the implanted side of the film through steps of electron-beam lithography (EBL), mask transfer, and reactive ion etching (RIE) with $O_2$ plasma. The diamond film is then thinned down from the backside with a shadow mask technique until the phononic structure is completely released and the desired thickness is attained. This is followed by stepwise thermal annealing up to a temperature of 1200 degrees for the formation of color centers. Soft $O_2$ plasma etching, wet chemical oxidation, and oxygen annealing are also used for the removal of damaged surface layers and for oxygen surface termination. Details of the fabrication procedures are given in the earlier work [21]. An optical image of cantilevers attached to squares in the phononic crystal lattice is shown in Fig. 1a.

A limitation of the above diamond nanofabrication approach is that the suspended structure retains the thickness variation of the initial diamond film, which results from the slicing and polishing of the thicker diamond film[21]. The thickness of the phononic structure can vary from 2.2 μm to 2.7 μm across the sample, leading to optical interference fringes as shown in the optical image of a phononic structure in Fig. 1b. Each color fringe corresponds to an estimated thickness variation of 110 nm. Since the phononic band gap depends on the thickness of the structure, an important question, which needs to be resolved experimentally, is whether robust protection of the mechanical modes by the phononic band gap can still be preserved in the presence of the inevitable thickness variations. Note that diamond nanofabrication techniques used in the earlier studies of optomechanical crystals and microdisks are not suitable for the fabrication of a few hundred μm sized or larger 2D structures[14, 18].

For the characterization of the out-of-plane mechanical modes in a cantilever, we have used the radiation pressure from a focused laser beam (with a wavelength near 1.55 μm) to excite the mechanical modes in the cantilever, for which the intensity of the laser beam is modulated sinusoidally in time with an electro-optic modulator. Mechanical vibrations of the cantilever are detected with the standard approach of laser interferometry, with the diamond phononic structure mounted on a sapphire wafer, as illustrated in Fig. 2a. For the laser interferometry, the reflection of a red laser beam (with a wavelength near 637 nm) from the cantilever and that from the sapphire wafer are coupled into a single mode fiber. The output from the fiber is detected with an amplified photodiode (New Focus 1801). A network analyzer (Keysight P5001A) is used for the measurement of the spectral response, more specifically scattering parameter $S_{21}$, of the relevant mechanical modes. For experiments at low temperature, the sample is mounted on a cold finger in an optical cryostat (Montana Instrument S50).

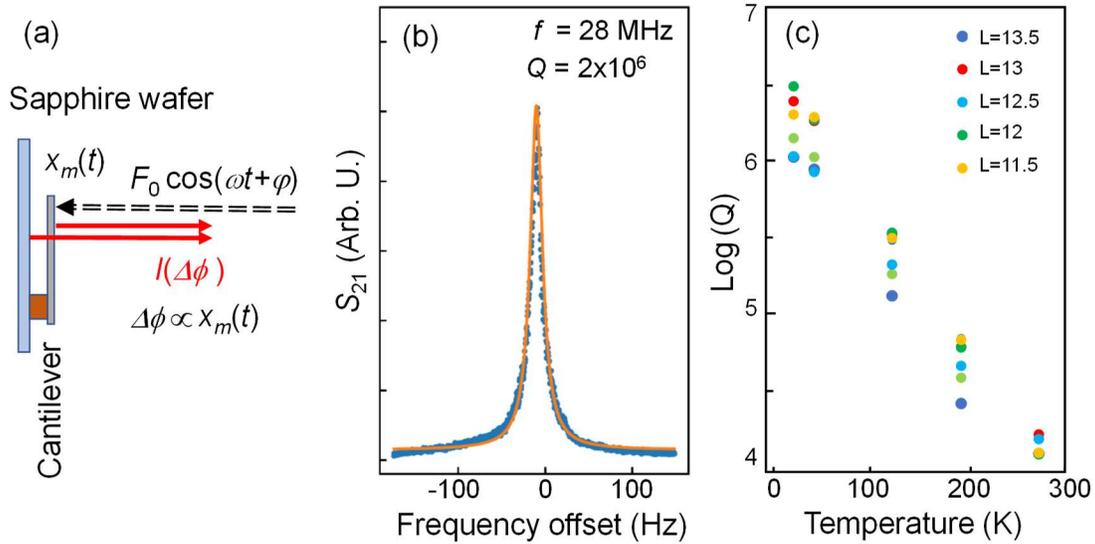

**Figure 2.** (a) Schematic depicting the excitation of out-of-plane modes in a cantilever with an intensity-modulated laser beam and the detection of the mechanical vibration with laser interferometry. (b) Spectral response of the fundamental out-of-plane mode in a cantilever obtained at T=8 K. The solid line is a least square fit to a Lorentzian. (c) Temperature dependence of $Q$ for fundamental out-of-plane modes in cantilevers with their lengths indicated in the figure.

Figure 2b shows an example of the spectral response of the fundamental out-of-plane mode in a cantilever obtained at 8 K. The spectral linewidth obtained (14 Hz) corresponds to $Q=2 \times 10^6$. Note that the Q-factors depend strongly on temperature. As shown in Fig. 2c, the Q-factors obtained increase by more than two orders of magnitude as the temperature decreases from room temperature to 8 K. Since we expect the clamping or structural loss to be independent or only weakly dependent on temperature, the strong temperature dependence shown in Fig. 2c indicates that Q-factors at temperatures as low as a few K are primarily limited by the materials loss. In this regard, additional experiments at temperatures as low as a few mK are still needed to determine the relevant mechanisms of mechanical damping in diamond nanomechanical resonators.

The variations in the structure thickness discussed earlier make it difficult to determine the precise spectral positions of the phononic band gap. For an estimation of the spectral variations of the band gap, we plot in Fig. 3 the calculated band edges of the lowest frequency phononic band gap as a function of the thickness of the phononic structure. The calculation shows that the variations of the lower band edge are on the order of 2 MHz and are considerable smaller than the variations of the higher band edge.

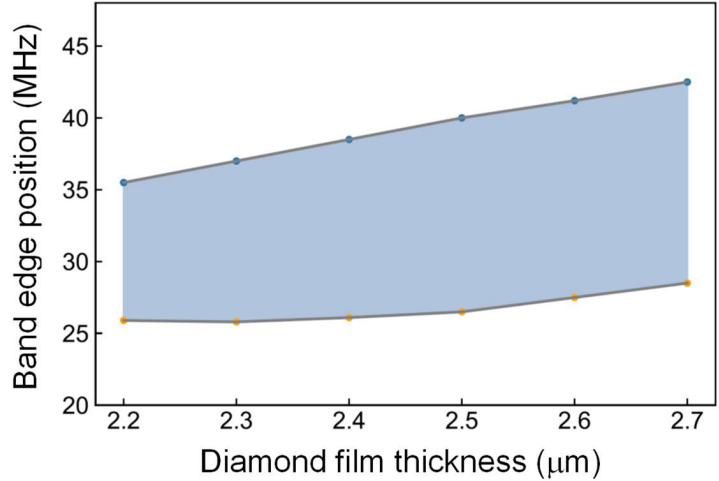

**Figure 3.** Calculated phononic band edge positions as a function of the diamond film thickness.

To demonstrate the protection of the mechanical modes by the phononic band gap even in the presence of the thickness variations, we have used a large number of cantilevers, which feature different lengths and thus different resonance frequencies. Figure 4 plots the Q-factors of the mechanical modes obtained from these cantilevers at temperatures near 8 K as a function of the corresponding resonance frequencies. For comparison, we also plot as shaded areas in Fig. 4 spectral ranges of the calculated band gaps for the square lattice, for which a structure thickness of 2.5 µm is used. Note that in addition to the fundamental out-of-plane modes, higher order modes can also be observed in the interferometry measurements. The Q-factors plotted near the second and third band gaps in Fig. 4 are obtained from the higher order mechanical modes.

As shown in Fig. 4, mechanical modes that are inside the band gaps of the phononic crystal feature Q-factors near or significantly above $10^6$. For many of these modes, a $f \cdot Q$ product greater than $10^{14}$ is observed. Note that it is much more difficult to excite and observe higher order mechanical modes with frequencies above the second band gap than the modes with much lower frequencies. For the high frequency modes, only modes with ultrahigh $Q$ can be observed. As such, we have only examined the high frequency modes in the third band gap in a few cantilevers.

Figure 4 also shows that as the mechanical resonance frequencies approach and then cross the lower frequency band edges, the Q-factors increase by nearly three orders of magnitude. Although we cannot precisely determine the band edge as discussed earlier, this large increase in the Q-factor occurs in a frequency range of only a few MHz, in agreement with the band gap calculations presented in Fig. 3.

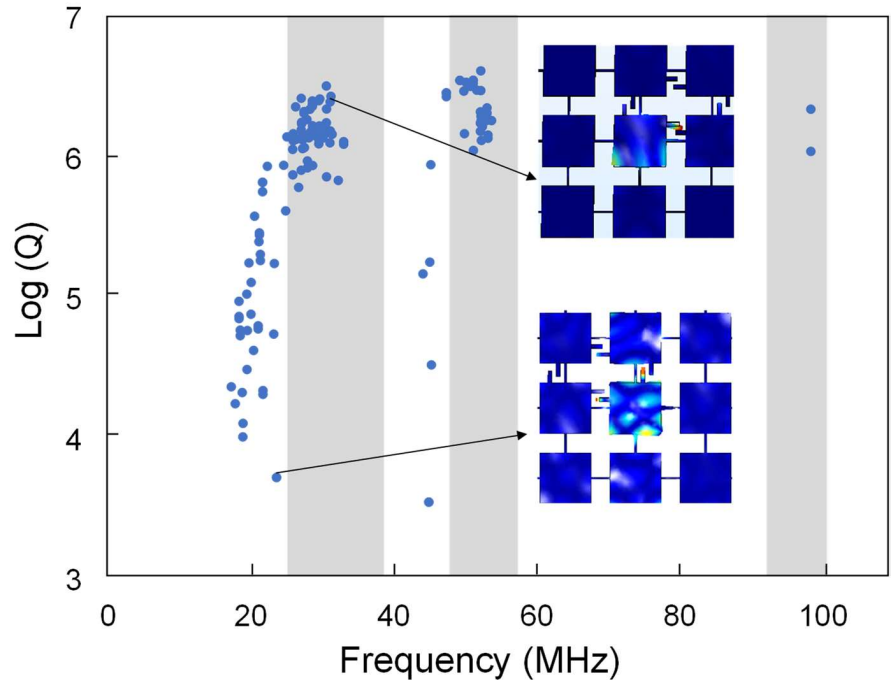

**Figure 4.** Q-factors obtained as a function of the resonance frequencies. The shaded areas indicate the calculated phononic band gaps of the square lattice with a diamond film thickness of 2.5 μm. The inset shows the COMSOL simulation of the mechanical displacement patterns of the two mechanical modes indicated in the figure. For the mode inside the gap, the displacement is confined to the square that hosts the cantilever. For the mode outside the gap, the displacement spreads out to neighboring squares.

The protection and isolation of the mechanical modes by the phononic band gap can also be seen from the COMSOL simulations. The inset in Fig. 4 shows, as an example, the calculated mechanical displacement patterns of two mechanical modes. For the high $Q$ mode inside the band gap, the displacement is confined within the square attached to the cantilever, indicating the isolation of the mechanical mode by the phononic band gap. In comparison, for the low $Q$ mode outside the band gap, the displacement spreads out to neighboring squares.

Diamond cantilevers embedded in a phononic crystal lattice can provide a promising system for pursuing spin-mechanics studies. For a NV center, the spin-mechanical coupling can take place through the strong excited-state strain coupling via a Raman transition[22, 23]. The coupling can be characterized by its cooperativity, a dimensionless parameter defined as $C = g^2/\gamma_m\gamma_s$, where $\gamma_m$ and $\gamma_s$ are the linewidths for the mechanical mode and the NV spin transition, respectively, and $g$ is the single-phonon spin-mechanical coupling rate. Our numerical estimations indicate that $C>1$ is achievable with the parameters of the cantilevers realized in this paper. For a specific estimate, we take $f$=28 MHz, $\gamma_m/2\pi$=14 Hz, $\gamma_s/2\pi$=0.6 MHz, the cantilever dimension to be (14, 4, 2.5) μm, the optical Rabi frequency and dipole detuning to be 0.2 GHz and 2 GHz, respectively, and the deformation potential to be 5 eV, which yields $C$=2.5. Furthermore, high $Q$ mechanical modes with relatively high resonance frequencies enabled by the phononic band gap protection make it possible to reach the resolved sideband regime for phonon-assisted optical as well as spin transitions, which can potentially allow processes such as phonon-mediated quantum state transfer and spin entanglement as well as cooling or amplification of mechanical motion via their coupling to color centers[6, 24].

In summary, we have achieved a $f \cdot Q$ product exceeding $10^{14}$ by using diamond cantilevers attached to a phononic square lattice. Our detailed experimental studies demonstrate the robust protection of mechanical modes by phononic band gaps and indicate that the mechanical Q-factors obtained at a few K are still limited by materials loss. Diamond nanomechanical resonators protected by a phononic band gap can open exciting opportunities for spin-mechanics studies.

**Funding sources**

This work is supported by the Air Force Office of Scientific Research (AFOSR) and by the National Science Foundation (NSF) under Grant Nos. 1719396, 1604167, and 2012524.

[*]Corresponding author email: hailin@uoregon.edu
[#]Present address: I.L.: US Naval Research Laboratory, Washington, DC 20375, USA

TOC graphics:

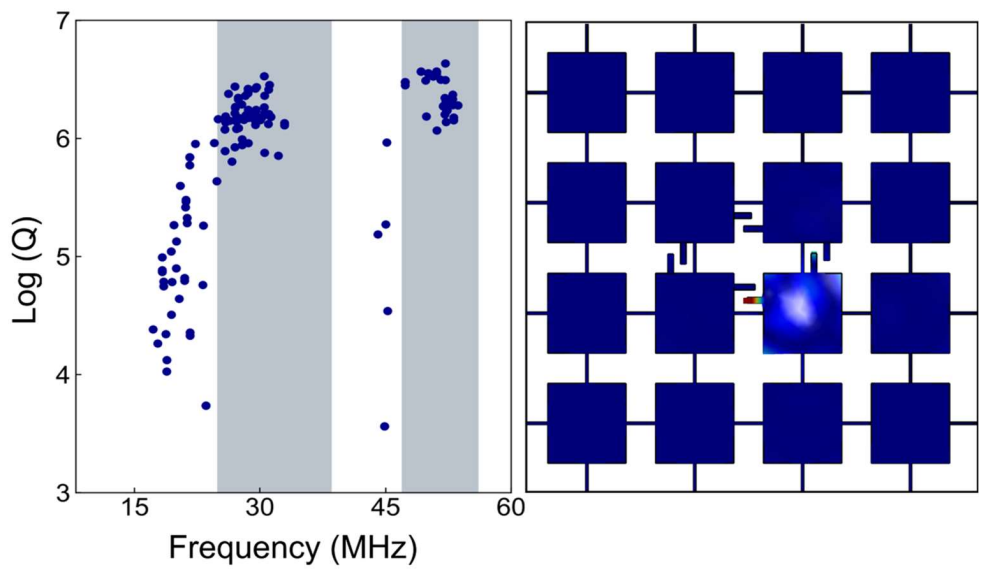